\documentclass{emulateapj}
\usepackage{epsfig,natbib}

\begin{document}

\slugcomment{ApJ in press } 
\shorttitle{Polarization of AU Mic} 
\shortauthors{Graham et al.}

\title{The Signature of Primordial Grain Growth in the
Polarized Light of the AU Mic Debris Disk }
\author{
James R. Graham\altaffilmark{1, 2, 3}, 
Paul G. Kalas\altaffilmark{1, 3}, \\
and \\
Brenda C. Matthews \altaffilmark{1, 4}
}
\affil{}

\altaffiltext{1}{Astronomy Department, University of California, 
Berkeley, CA 94720-3411, U.S.A.}
\altaffiltext{2}{jrg@berkeley.edu}
\altaffiltext{3}{Center for Adaptive Optics, 
University of California, Santa Cruz, CA 95064, U.S.A. }
\altaffiltext{4}{Herzberg Institute of Astrophysics,
National Research Council of Canada, 5701 West Saanich Road
Victoria, BC, Canada V9E 2E7}

\begin{abstract}
We have used the coronagraphic mode of the Advanced Camera for Surveys
aboard the {\it Hubble Space Telescope} to make the first polarization
maps of the debris disk surrounding the nearby M star AU Microscopii
(GJ 803). The linear polarization of the scattered light from the disk
is unambiguously detected.  We find that the degree of polarization
for the disk rises monotonically from about 0.05 to 0.35 between
projected radii of 20 and 50 AU.  Polarized light is detectable out to
about 80~AU, where the fractional polarization reaches a maximum
observed value of $0.41\pm 0.02$.  Polarization vectors are oriented
perpendicular to the disk mid-plane, indicating that polarization
originates from single scattering in an optically thin dust disk where
the albedo is dominated by small ($x = 2\pi a/\lambda \lesssim 1$)
particles.  We use simple, optically thin disk models to infer the
spatial structure of the disk and the scattering properties of the
constituent grains by simultaneously fitting the surface brightness
profile and the degree of linear polarization.  The best fit models
require that the dust grains exhibit high maximum linear polarization
and strong forward scattering.
The inner disk ($<$40--50~AU) is depleted of micron-sized dust by a
factor of more than 300, which means that the disk is collision
dominated, i.e., grains that are dragged inward by corpuscular and
Poynting-Robertson drag undergo a destructive collision.
While the inferred optical properties are covariant with the radial
distribution of dust, the only acceptable models have $p_{max} \ge
0.50$ and $g \ge 0.7$.  These constraints cannot be met by spherical
grains composed of conventional materials.  A Mie scattering analysis
implicates grains where the real part of the refractive index is about
1.03, which is a signature of highly porous (91--94\%) media.  More
reliable methods for calculating the scattering properties of
aggregates confirm that the observed values of $p_{max}$ and $g$ can
be matched by high porosity, micron-sized clusters of small particles.
In the inner Solar System, porous particles form naturally in cometary
dust, where the sublimation of ices leaves a ``bird's nest'' of
refractory organic and silicate material.  In AU Mic, the grain
porosity may be primordial, because the dust ``birth ring'' lies
beyond the ice sublimation point.  The observed porosities span the
range of values implied by recent laboratory studies of particle
coagulation in the proto-solar nebula by ballistic cluster-cluster
aggregation.  To avoid compactification, the upper size limit for the
parent bodies is in the decimeter range, in agreement with theoretical
predictions based on collisional lifetime arguments.  Consequently, AU
Mic may exhibit the signature of the primordial agglomeration process
whereby interstellar grains first assembled to form macroscopic
objects.

\end{abstract}

\keywords{polarization---dust---stars: individual(\objectname{GJ 803;
AU Mic})---planetary systems: formation--- circumstellar matter}

\section{Introduction}

Approximately 15\% of nearby main sequence stars manifest infrared
excess due to the reprocessing of stellar radiation by dust grains in
a circumstellar disk \citep{1984ApJ...278L..23A, 1993prpl.conf.1253B}.
These systems are known as ``debris disks'' because the lifetime of
dust is orders of magnitude shorter than the stellar age, suggesting a
continuous supply of fresh grains released from larger, undetected
parent bodies.  In the solar system, interplanetary dust particles are
resupplied by the collisional erosion of asteroids and the sublimation
of comets.  Cometary dust grains retain the history of their
interstellar origin, but are imprinted with structures that speak to
their incorporation into larger bodies and subsequent return to
interplanetary space \citep{1990ApJ...361..260G}.  Therefore, cometary
dust provides our closest link to the particle coagulation processes
that occurred during the earliest phases of planet building in the
Solar System.  Although debris-disk dust must be highly processed and
modified during incorporation into, and subsequent attrition of, large
bodies, these particles are our primary source of information
regarding the growth of assembly of planetesimals in exoplanetary
systems.  The laboratory study of low-velocity dust interactions
thought to be characteristic of conditions in the early solar nebula
suggests that particles grow under ballistic cluster-cluster
aggregation into fractal assemblies
\citep{1998Icar..132..125W}. These bodies suffer restructuring when
the aggregate diameters exceed a few centimeters. Smaller bodies are
not expected to be subjected to impact compaction
\citep{2000Icar..143..138B}.  As these clusters have unique optical
properties, the study of debris disks provides experimental validation
of our ideas regarding growth of solid bodies from interstellar grains
to macroscopic objects \citep{2006A&A...449.1243K}.

At visible wavelengths debris-disk dust can be detected in scattered
light, analogous to the Zodiacal light in our Solar System.  As in the
Solar System, the polarization state of this scattered light is a key
diagnostic of the grain properties.  In an optically thin disk, where
single scattering dominates, a high degree of linear polarization is
expected, with a characteristic orientation that is symmetric
(concentric or radial) about the illuminating source
\citep[e.g.,][]{2003pid..book.....K}.  The observed degree of
polarization is an important indicator of the size, shape,
composition, physical structure and alignment of individual grains and
their distribution along the line of sight.  Single particle
scattering can be described by the matrix elements of the complex
amplitude scattering function, $S$, which depend on the scattering
angle \citep{1981lssp.book.....V}.  The angular dependence of $S$ is a
key clue to the nature of the particles. For example, the degree of
forward scattering increases with particle size from the Rayleigh
limit ($x=2\pi a/\lambda \ll 1$), where the scattering asymmetry
parameter, $g = \langle \cos \theta \rangle \simeq 0$, to $g\simeq 1$
as $x$ approaches unity.  Observations of spatially resolved disks can
directly determine the azimuthal asymmetry due to asymmetric
scattering \citep[e.g.,][]{2005Natur.435.1067K}.  Because debris disks
are optically faint relative to their illuminating star, there is a
strong selection effect that favors the discovery and facilitates the
study of high surface brightness, edge-on systems, e.g., $\beta$ Pic,
HD 32297, HD 139664, or AU Mic.

In an edge-on system, the azimuthal variation of surface brightness
around the star is unobservable, and thus grain scattering properties
versus phase angle cannot be measured directly.  Each line of sight
through the disk includes a range of scattering angles.  Therefore,
both the asymmetry parameter and the radial variation of dust opacity
determine the disk surface brightness as a function of angle on the
sky.  Unless there is prior information regarding $g$ or the dust
distribution these properties cannot be uniquely disentangled from
measurements of the radial surface brightness distribution.  Inclusion
of further constraints, such as the spectral energy distribution
\citep[e.g.,][]{2005astro.ph.10527S}, can break this degeneracy.
Polarization data can also play this role.  In general, the matrix
elements of {\sf S} have different angular dependences. Therefore,
observation of the intensity, which depends on $|S_1|^2+|S_2|^2$, and
the polarization state of scattered light, which depends on
$|S_1|^2-|S_2|^2$, can be used to recover this otherwise lost
information.

Here we present the first optical polarization study of the AU Mic
debris disk.  AU Mic is a nearby (9.9~pc) dM1e star with Galactic
space velocities that suggest a common origin with $\beta$ Pic.
\citep{1999ApJ...520L.123B}.  The discovery of scattered visible light
from a near-edge-on debris disk around AU Mic
\citep{2004Sci...303.1990K} supports the picture of $\beta$ Pic, AU
Mic and nearly 20 other stars as a coeval group with age
12$^{+8}_{-4}$~Myr \citep{2001ApJ...562L..87Z}.  The $\beta$ Pic
debris disk has been studied extensively, including the finding of
polarization along the mid-plane varying between 0.12 and 0.21 in
optical data \citep{ 1991MNRAS.252P..50G, 1995Ap&SS.224..395W}.  These
measurements have been interpreted by several authors
\citep[e.g.,][]{1997AREPS..25..175A, 1999A&A...352..508V,
2000ApJ...539..424K}.  In recent ground based observations,
polarization in the $K$ band has been detected
\citep{2006ApJ...641.1172T}.

Despite the large difference in stellar mass,
\citet{2005astro.ph.10527S} and \citet{2006astro.ph..4313A} argue that
the radiation pressure force that quickly expels small $\beta$ Pic
grains has a counterpart around AU Mic in the form of stellar wind.
To first order this explains the initial finding that the disk
mid-plane surface brightness distribution is nearly identical for both
debris disks \citep{2004Sci...303.1990K}.  However, the blue color
gradient for AU Mic's mid-plane \citep{2005ApJ...622..451M,
2005AJ....129.1008K, fitz06} contrasts against the red color of
$\beta$ Pic's mid-plane \citep{2006AJ....131.3109G}.  This distinct
color difference between the two disks points to a significant
divergence in either the grain composition, grain size distribution,
or minimum grain size.  Polarization observations of AU Mic are
therefore valuable in constraining these grain properties
and identifying the fundamental differences between the two disks.

Our purpose is to present Advanced Camera for Surveys (ACS)/High
Resolution Camera (HRC) polarimetry of the AU Mic system and to shed
some initial light on the disk structure and grain optical properties
that are emphasized by the detection and measurement of the disk in
polarized light.  The fidelity of any model increases with the
observational challenges presented to it, thus it must fit the near-IR
and optical emission \citep{2004Sci...303.1990K, 2004Sci...305.1442L,
2005ApJ...622..451M, fitz06} and the IR to sub-mm spectral energy
distribution \citep{2004ApJ...608..526L, 2005ApJ...634.1372C}.  We
defer this synthesis to a later study.

Section \ref{observations} outlines the observations, point spread
function (PSF) subtraction, calibration of the Stokes parameters, and
correction for instrumental and interstellar polarization.  Section
\ref{results} reports the appearance of the AU Mic disk in polarized
light.  Section \ref{onedprofiles} provides a qualitative description
of the extracted one-dimensional surface brightness profile and degree
of linear polarization, compares AU Mic with $\beta$ Pic, and draws
some preliminary conclusions regarding the radial distribution of the
dust.  We describe quantitative analysis using optically-thin edge-on
disk models in \S \ref{diskmodels} and develop a method that
simultaneously fits the observed surface brightness and fractional
polarization to three dust models: a semi-empirical Henyey-Greenstein
model, a Zodiacal dust model, and a Mie model.  The successes and
failures of these models are addressed in \S \ref{discussion} and the
evidence for porous grains is examined.  Our conclusions are
summarized in \S \ref{summary}.

\section{Observations, Data Reduction and Calibration}
\label{observations}

Coronagraphic observations of AU Mic (GJ 803, HD 197481) were made on
2004 August 01, using the $1\farcs8$ diameter ($\simeq 64 \lambda /D$
at $V$) occulting spot on the ACS/HRC aboard {\it Hubble Space
Telescope} (HST).  The occulting spot is located in the aberrated beam
from HST, before corrective optics, and intercepts about 88\% of the
on-axis light \citep{2003SPIE.4854...81F}.  The F606W broadband
filter, comparable to the Johnson-Cousins $V$ band, was chosen for all
observations.  AU Mic was observed during two orbits with the
spacecraft roll angle offset by 10.1 degrees between orbits.  Three
exposures of 240 seconds were obtained through each of the three
polarizer elements (POL0V, POL60V, POL120V).  A third orbit was
devoted to observing a PSF reference star using the same filter
combinations.  Together, PSF-subtracted, coronagraphic data render an
improvement factor exceeding one hundred in contrast relative to
direct imaging.  The PSF star (GJ 784; M0V
\citep{1957MNRAS.117..534E}) was chosen on the basis of similar
brightness to AU Mic, close spectral type match, and proximity on the
sky.  Data reduction included the standard pipeline processing from
the HST archive that produces bias-subtracted and flat-fielded image
files.  Images were additionally processed using the recommended
spotflat and pixel area map.  Differencing frames within an orbit and
between orbits was used to register all images.

PSF subtraction at each POLV filter was achieved by subtracting the
PSF reference star from AU Mic to produce a residual image that gives
a mean radial profile equal to zero intensity in directions
perpendicular to the mid-plane. PSF subtraction was implemented before
construction of the Stokes parameter images because the POLV filters
introduce filter-specific artifacts.  The polarizer filters contribute
a weak geometric distortion which rises to about 0.3 pixels near the
edges of the HRC.  This is caused by a weak positive lens in the
polarizers, which is needed to maintain proper focus when multiple
filters are in the beam. In addition, the visible polarizer has a weak
ripple structure which is related to manufacture of its Polaroid
material; this contributes an additional $\pm$ 0.3 pixel distortion
with a complex structure \citep[][]{2004biretta-10,2004kozhurina}.
All these geometric effects are correctable, but astrometry obtained
with the POLV filters will likely have reduced accuracy due to
residual errors and imperfect corrections.

\begin{figure}
\plotone{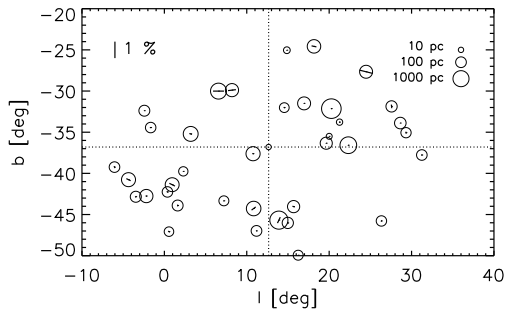}
\caption{ Interstellar polarization for stars in the neighborhood ($<$
25 degrees) of AU Mic \citep{2000AJ....119..923H}.  Stars are plotted
in Galactic coordinates, centered at the location of AU Mic
(intersecting dotted lines).  The length of the line designates the
degree of linear polarization, and the diameter of the circle is
proportional to the logarithm of the distance.  The vertical tick mark
labeled 1\% gives the degree of polarization scale.  The median
measured polarization is 0.1\% in this direction suggesting that the
interstellar polarization of AU Mic is small and can be neglected. }
\label{istelpp}
\end{figure}

\subsection{Calibration of the Stokes Parameters}
\label{stokescal}

Imaging polarimetry with ACS is described and characterized by
\citet{2004biretta-09} and \citet{2006pavlovsky}.  The ACS/HRC
polarimeter is implemented as three analyzers installed in a filter
wheel at nominal angles of 0, 60 and 120 degrees denoted POLV0,
POLV60, and POLV120, respectively.  Mueller matrix algebra can be used
to show that the observed intensity, $I_\theta$, through a perfect
analyzer rotated by an angle $\theta$ is
\begin{equation}
I_\theta = \frac{1}{2}
\left(I + Q\cos2\theta + 
U \sin 2\theta \right) , 
\label{modulator}
\end{equation}
where $I$, $Q$ and $U$ are the 
Stokes parameters \citep{1960ratr.book.....C}. 
This arrangement is insensitive to the 
circular polarization, $V$, which we 
assume henceforth to be negligible. 
Assuming that the three POLV filters are ideal,
and ignoring instrumental polarization, we can solve the set
of resultant simultaneous equations to show
\begin{eqnarray}
I & = & \frac{2}{3}\left( I_0 + I_{60} + I_{120}  \right) \\
Q & = & \frac{2}{3}\left( 2I_0 - I_{60} - I_{120}  \right) \\
U & = & \frac{2}{\surd{3}}\left( I_{60} - I_{120}  \right) .
\label{stokesconversion}
\end{eqnarray}
We measure instrumental angles counter-clockwise from the $+Q$ axis,
which we take to be defined by the orientation of the POLV0 filter.
We quote this elementary result because \citet{2004biretta-09} and
\citet{2006pavlovsky} each use different sign conventions for angles,
resulting in different expressions for $U$; our choice is consistent
with the latter.  The Stokes parameters are then projected onto an
astronomical coordinate system using the Mueller matrix to rotate from
the spacecraft reference frame. The degree of linear polarization,
which is occasionally referred to as the fractional polarization, is
defined as $p = (Q^2 + U^2)^{1/2}/I$. The position angle of the
electric field is $\psi = (1/2) \arctan(U/Q)$.

The polarizing efficiency of the POLV filters is high: $(T_s -
T_p)/(T_s + T_p) > 0.999$ at 600~nm, where $T$ is the transmission for
the $s$ and $p$ linear polarization states, respectively, and the
orientation of each POLV filter is within a degree of its nominal
values \citep{2004biretta-09,2004biretta-10}.  Nonetheless, the ACS is
not an ideal polarimeter.  The HRC employs three non-normal
reflections and a tilted detector that combine to yield an
instrumental polarization in F606W of 0.063 at a position angle of
$\psi = -87$ degrees\footnote{Using our sign convention.}
\citep{2004biretta-09}.

\clearpage
\onecolumngrid

\begin{figure}
\plotone{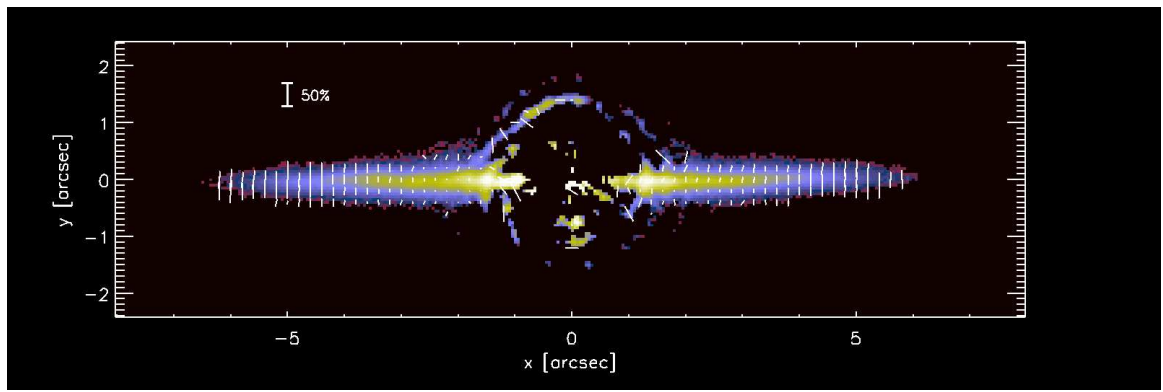}
\caption{A pseudocolor representation of the AU Mic debris disk in
Stokes $I$ measured with ACS/HRC in F606W ($\lambda_c$= 590 nm,
$\Delta\lambda$ = 230 nm).  Over-plotted are ticks that indicate the
orientation of the electric field. The length of the tick is
proportional to the degree of linear polarization.  A 50\%
polarization tick is indicated.  The E-field vectors are derived from
Stokes parameters that have been binned $\times 8$ into $0\farcs2$
pixels prior to calculating the degree of polarization.  The binned
vectors are fully independent.  The degree of polarization rises
smoothly from about 5\% close to the star up to approximately a peak
linear polarization of 40\%.  The high degree of polarization, and the
orientation of the electric vector perpendicular to the disk plane are
indicative of small-particle scattering in an optically thin disk.
Data within a radius of about $ 1\farcs0$ suffer from significant
systematic errors due to imperfect PSF subtraction. }
\label{polvplot}
\end{figure}

\twocolumngrid

We remove the instrumental polarization from our data by using the
correction factors, which are derived by \citet{2006pavlovsky} from
observations of polarization standard stars.  These corrections are
applied to the observed count rate in each of the three polarizers
before computing the Stokes parameters.  The systematic error in the
degree of linear polarization for a weakly polarized source is about
$\pm$ 0.01; the resultant degree of linear polarization will have a
{\em fractional} systematic error of about 10\% for highly polarized
($p>$ 0.20) sources. For example, the systematic error for $p$ = 0.05 is
0.01, but the systematic error for $p$ = 0.50 is 0.05.  The systematic
error in $\psi$ is about 3 degrees.  Because we combine two
observations at different roll angles, our systematic errors are
somewhat reduced relative to these values.

The $V$-band polarization standards BD +64$^\circ$106 ($p=569\pm
4\times 10^{-4}$; $\psi = 96.6 \pm 0.2$ degrees) and GD 319 ($p=9\pm 9
\times 10^{-4}$) \citep{1992AJ....104.1563S} were observed with the
HRC, F606W, and the POLV filter set, as part of the ACS polarization
calibration program (HST Proposal IDs 9586 \& 9661).  We established
the correctness of our implementation of the Stokes parameter
calibration procedure by measuring the degree of polarization and PA
of these stars, comparing with the ground-based results, and
confirming agreement within the statistical errors.  The polarization
calibration targets are weakly polarized point sources, whereas the AU
Mic disk is spatially extended.  To gain experience and confidence
with imaging polarimetry we also analyzed the ACS Wide Field Camera
F606W/POLV observations of the highly polarized Crab Pulsar
synchrotron nebula (Proposal ID 9787).

\subsection{Interstellar Polarization}

In additional to correction for instrumental polarization observations
should also be corrected for interstellar polarization.  AU Mic is
nearby, at high Galactic latitude and has minimal color excess,
suggesting that polarization due to interstellar dust grains is likely
to be negligible. Although the polarization of AU Mic itself has not
been detected in integrated light \citep{1981ApJ...247.1013P}, we can
inspect the interstellar polarization of adjacent stars
(Fig. \ref{istelpp}).  This figure shows the observed degree of linear
polarization for stars within an angular radius of 25 degrees of AU
Mic from \citet{2000AJ....119..923H}. The median measured polarization
is $1\times 10^{-3}$ in this direction on the sky and the highest
observed value is $8\times 10^{-3}$, and this is for a star
considerably more distant (260 pc) than AU Mic. We therefore neglect
any interstellar polarization in the subsequent discussion. AU Mic is
a flare star, and may exhibit flare-generated particle beam polarized
light \citep[][]{1987ApJ...312..822K,1994A&A...286..194S}.

\section{Results}
\label{results}

\subsection{Linear Polarization of the AU Mic Disk}
\label{linpol}

Stokes $I$, $Q$ and $U$ images were created from the PSF-subtracted
POLV0, POLV60 and POLV120 images as outlined in \S \ref{stokescal}.
The corresponding degree of linear polarization and the orientation of
the electric field are represented in Figure \ref{polvplot}.  Because
the Stokes parameters are formed from a linear combination of the
observed flux, their statistical properties are simple. In contrast,
the degree of polarization is a positive definite quantity, and
therefore biased. Consequently, the polarization information that we
display in Figure \ref{polvplot} has been derived from well binned
Stokes images prior to computation of $p$ and $\psi$ to ensure that
this figure gives a reliable impression of the results.

\begin{figure}
\plotone{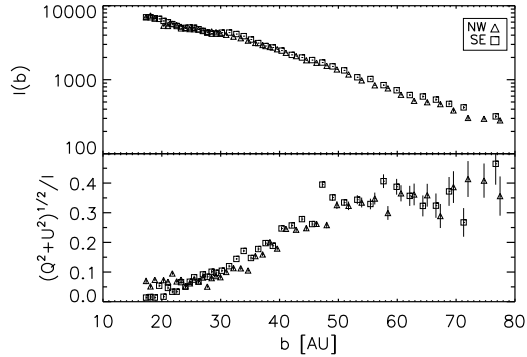}
\caption{One dimensional Surface brightness in Stokes $I$ (top) and
degree of linear polarization (bottom) as a function of projected
radius, $b$, for AU Mic.  The surface brightness is in units of
detected photoelectrons per $0\farcs025$ (0.25 AU) wide pixel column.
Errors represent statistical errors only. Systematic errors are
included in Figure \ref{hggzodi}.  }
\label{onediandp}
\end{figure}

\begin{figure}
\epsscale{0.5}
\plotone{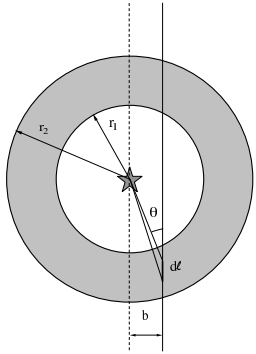}
\epsscale{1}
\caption{Model disk geometry. The surface brightness at impact
parameter $b$ is evaluated along the line of sight, $\ell$.  The
scattering cross section is a function of the scattering angle,
$\theta = \arcsin(b/r)$.  It is convenient to make the change of
variable in Eq. (\ref{sb_integral}) to $\theta$, in which case the
limits of integration become $\arcsin(b/r_2)$ and $\pi -
\arcsin(b/r_2)$.  }
\label{geometry}
\end{figure}

Figure \ref{polvplot} shows that the degree of polarization increases
monotonically with distance from the star from about 0.05 to 0.40 (see
also Fig.  \ref{onediandp}).  Everywhere, the electric field is
consistent with an orientation perpendicular to the disk.  These two
results are qualitatively in accord with the expected signature of
scattering by small spherical particles with $x \lesssim 1$
\citep{2003pid..book.....K}.  It is likely that the variation of
polarization with distance from the star occurs because, along a given
line of sight, a range of scattering angles contributes to the
observed intensity. The degree of linear polarization typically peaks
at scattering angles close to $\pi/2$, and polarization is always zero
in the forward and backward directions.  Thus, the peak polarization
signal is diluted by light arising from more acute and more oblique
scattering events.  We expect to see the peak polarization at the
outer edge of the disk, where only right-angle scattering contributes.
If the disk is devoid of dust within some inner boundary $r_1$, then
for impact parameters $b < r_1$ scattering events with angles between
$\arcsin (b/r_1)$ and $\pi - \arcsin (b/r_1)$ are absent, and the
degree of polarization is reduced even further (see Figure
\ref{geometry}).  For an optically thin disk, where single,
small-particle scattering dominates, the electric field is oriented,
as observed here, perpendicular to the plane containing the star, the
dust grain and the observer. For intermediate size ($x \gtrsim 1$)
spherical particles the plane of polarization can flip by $\pi/2$ at
certain scattering angles so that the electric field is oriented
parallel to the scattering plane \citep{2003pid..book.....K}, which is
clearly not the case here.  Moreover, particles composed of
conventional astrophysical grain materials show large amplitude
oscillatory behavior in $|S_1|^2-|S_2|^2$ with scattering angle, with
angular period $\delta \theta \simeq \lambda/2a$.  Thus any line of
sight that comprises emission from a range of scattering angles,
$\Delta \theta$, such that the particle size satisfies $\Delta \theta
/ \delta \theta \gg 1 $, will tend to exhibit weak linear
polarization.


\subsection{One-dimensional Profiles}
\label{onedprofiles}

\begin{figure}
\plotone{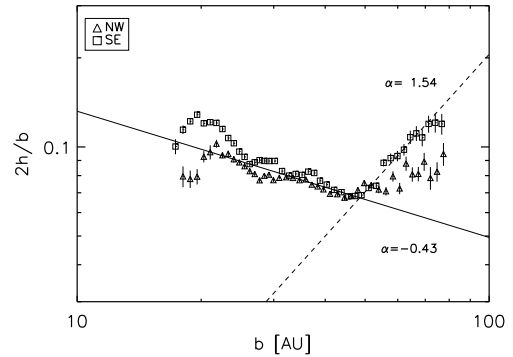}
\caption{Vertical thickness of the disk (in units of $b$) derived from
F606W Stokes $I$ as a function of projected separation, $b$.  The
projected FWHM of the disk is measured by fitting a Cauchy function
(Eq. (\ref{cauchy}), where $2h$ is the FWHM.  The strong variation of
disk thickness means that photometry in a fixed aperture is not an
effective means for extracting the 1-d surface brightness profile.
The vertically integrated surface brightness profile shown in
Fig. \ref{onediandp} is simply $C(b)$.  The disk thickness varies with
projected separation, and shows two distinct regimes: within 50 AU $h
\sim b^{1/2}$ and beyond 50 AU $h \sim b^{5/2}$.  The thin lines show
robust least-squares fits of the form $h/b \sim b^\alpha$ for $20 <
b/AU < 50$ (solid) and $b \ge 50$ (dashed).  }
\label{scalehplot}
\end{figure}

We have measured one-dimensional surface brightness profiles along the
disk in $I$, $Q$ and $U$ for comparison with simple disk models.  This
photometry was extracted optimally using column-by-column fitting to
the vertical surface brightness profiles. This approach is
advantageous because the mid-plane location and projected disk
thickness vary significantly with impact parameter
\citep{2005AJ....129.1008K}.  Figure \ref{scalehplot} shows the
full-width at half maximum (FWHM) thickness of the disk (in units of
the impact parameter, $b$), as a function of $b$. Between projected
radii of 25 and 50~AU the FWHM grows relatively slowly as $\sim
b^{1/2}$, then beyond 50~AU the thickness increases more rapidly as
$\sim b^{5/2}$.  Because of this variation, it is unsatisfactory to
simply measure the disk signal in an aperture of fixed height: a
varying fraction of the emission is missing from a small aperture,
while excess noise contaminates large apertures.  Neither an
exponential nor a Gaussian describe the vertical profile adequately.
However a Cauchy distribution
\begin{equation}
c(b,z) = C(b) \; \frac{h}{\pi [h^2 + (z-z_0)^2]},
\label{cauchy}
\end{equation}
with FWHM, $2h$, provides an excellent fit at all impact parameters,
where $z$ is the coordinate perpendicular to the disk plane, and $z_0$
is the location of the mid-plane.  The 1-d surface brightness profile
is then simply the values $C(b) = \int c(b,z)\; dz$ derived from this
fit.  The data displayed in Figure \ref{scalehplot} are derived from
fitting the Stokes $I$ image.  Because the surface brightness declines
with radial distance, we have binned the vertical profile in
increasingly wide blocks with distance from the star. The data are
binned into three-pixel ($0\farcs075$) wide columns for $r \le 30$~AU
increasing to nine-pixel ($0\farcs225$) wide columns for $50 < r/{\rm
AU} \le 70$~AU.  This binning reduces the uncertainties at the outer
edge of the disk at the expense of lower angular resolution.  The
signal-to-noise is greater in Stokes $I$ than in $Q$ or
$U$. Therefore, we assume that the disk thickness and mid-plane
location does not vary with the polarization state and use the results
from fitting Stokes $I$ to constrain the $Q$ and $U$ fits.  The
resultant Stokes $I$ 1-d surface brightness profile and degree of
linear polarization is shown in Figure \ref{onediandp}.

The polarization measurements of $\beta$ Pic's debris disk provide a
natural point of reference for comparison with our results.  A
qualitative consideration of the factors described in \S \ref{linpol}
suggests that these two disks have different polarization signatures
because they have different radial dust distributions, and the two
disks are measured on different scales.  The $R$-band polarization of
$\beta$ Pic's disk is observed to range from 0.12 to 0.17
\citep{1991MNRAS.252P..50G, 1995Ap&SS.224..395W}; the degree of
polarization shows a weak gradient, increasing outward between 200 and
600 AU.  In contrast, polarization of the AU Mic disk rises quickly by
a factor of about five between projected radii of 20 and 50 AU.  A
direct comparison of AU Mic with $\beta$ Pic is not possible because
the spatial scales probed do not overlap---partly because $\beta$ Pic
at 19.3 pc is almost twice as distant as AU Mic, and partly because
the $\beta$ Pic measurements are derived from seeing limited
observations.  Nonetheless, the differences suggest that the rapid
rise of the linear polarization of AU Mic's disk between 20 and 50 AU
occurs because these lines of sight intersect a central hole where
scattering angles $\simeq \pi/2$ are absent. In $\beta$ Pic this steep
rise in unobserved, because the dust depleted zone lies too close to
the star to be readily observable from the ground.  
Recent near-IR adaptive optics data that probe $\beta$ Pic's
debris disk on scales of 60--120 AU are consistent with a 120 AU inner
hole radius \citep{2006ApJ...641.1172T}.  The visible extent of the
$\beta$ Pic disk is at least 1800 AU \citep{2001MNRAS.323..402L}.
Evidently, $r_1/r_2$ is small ($\simeq 0.1$) for $\beta$ Pic, and
assuming that the optical properties of grains are homogeneous across
this disk then the polarization should continue to increase gradually
with increasing impact parameter out to the outer edge, located 
at $r_2$.

This explanation, which invokes only geometric factors to explain the
difference between AU Mic and $\beta$ Pic, is incomplete on two
counts.  First, the outer radius cannot be $\simeq 100$ AU, because
the AU Mic disk is traced out to 210 AU \citep{2004Sci...303.1990K}.
Second, the peak detected polarization beyond 50 AU exceeds 0.30,
which exceeds the peak linear polarization of the dust invoked to
explain the $\beta$ Pic measurements.  Taken together these two
observations imply that the peak linear polarization of an individual
scatterer in the AU Mic disk must exceed 0.40. However, a quantitative
comparison in \S \ref{diskmodels} shows that our inferences about the
relative scales of these two disks contain a grain of truth.

\section{Disk Models}
\label{diskmodels}

The linear polarization of $\beta$ Pic's disk can be explained by
assuming a radial, power-law opacity distribution and the optical
properties of Solar System dust grains, examples of which include the
Zodiacal-light grains, interplanetary dust particles (IDPs), and
cometary dust \citep{1997AREPS..25..175A}.  It is therefore useful to
enquire whether the polarization signature measured in \S
\ref{onedprofiles} can be described by such a model, and whether the
differences between AU Mic and $\beta$ Pic (\S \ref{linpol}) can be
attributed solely to different radial opacity distributions.

Suppose that the surface density of dust follows a power law, and the
grain properties are uniform throughout the disk, such that the
vertical optical depth to scattering presented by the grains is
$\tau_\perp(r) = \int_z n(r,z) \pi a^2 Q_{sca} \; dz = \tau_{\perp
,1}(r_1/r)^{\beta}$.  The differential scattering cross section per
unit solid angle of each dust grain is $\pi a^2 Q_{sca} s_i(\theta)$,
where $a$ is the geometric grain radius and $Q_{sca}$ is the
scattering efficiency.  We use $i$ to denote either the $s$ or $p$
polarization states (with respect to the scattering plane), and
$\theta$ indicates the dependence of the cross-section on the
scattering angle. The phase function, $s_i$, is normalized so that
$\int s_i\; d\Omega=1$.  For anisotropic particles, the cross-sections
are functions of $\theta$ and $\phi$.  We assume isotropy, and ignore
the azimuthal dependence, i.e., the grains are spherical, or randomly
oriented.

The observed one-dimensional surface brightness
(cf. Eq. (\ref{cauchy})) of an optically thin disk is expressed by an
integral along the line of sight, $\ell$,
\begin{equation}
C_i (b) = L_\nu \int_{-(r_2^2-b^2)^{1/2}}^{(r_2^2-b^2)^{1/2}} 
\frac{\tau_\perp(r)}{4 \pi r^2}s_i(\theta) \; d\ell,
\label{sb_integral}
\end{equation}
where $L_\nu$ is the monochromatic luminosity of the star at the
observing frequency, $r$ is the radial coordinate, and $r_1$ and $r_2$
are the inner and outer radii of the disk (see Figure \ref{geometry}).
This integral can be rewritten by change of variable to the scattering
angle using $r^2\; d\theta = -b\;d\ell$ and setting the limits of
integration to $\theta_2 = \arcsin (b/r_2)$ and $\pi - \theta_2$.
Evaluation of the surface brightness for a disk with an inner hole,
radius $r_1$, is convenient using this form, because the integral can
be written as the sum of two contributions from scattering angles
between $\theta_2$ and $\theta_1$ and between $\pi - \theta_1$ and
$\pi - \theta_2$.  The Stokes parameters are then calculated according
to
\begin{eqnarray}
I(b) & = & C_p (b) + C_s (b) ,\nonumber  \\
Q(b) & = & C_p (b) - C_s (b) ,\nonumber  \\
U(b) & = & 0.
\label{sandptostokes}
\end{eqnarray}
According to this convention, the
scattering plane (the disk)
defines the $+Q$ axis. 
If the grains are asymmetric and their orientations are not random,
or the disk is not exactly edge on
then $U\neq 0$. However, there is no evidence from our current
observations that this assumption is violated.  As the disk is
optically thin, multiple dust components can be represented by summing
the contribution from different grain populations; such components may
include grains of various sizes or composition.

\subsection{A Henyey-Greenstein Model}
\label{hgmodel}

We begin with a simple grain scattering model that illustrates both
the feasibility of simultaneous fitting of the surface brightness and
degree of polarization and the nature of the resultant constraints.
Although any phase function can be written as a sum of Legendre
polynomials, our goal is to construct a model with the minimum number
of free parameters. Therefore, we adopt the empirical
Henyey-Greenstein function as a convenient approximation to scattering
by small particles \citep{1941ApJ....93...70H}.  We assume that the
azimuthal dependence of polarization obeys a Rayleigh law, with peak
linear polarization parameterized by $p_{max}\le 1$.  The
corresponding elements of the intensity scattering matrix are given by
\begin{equation}
\frac{1}{2}\left(|S_1|^2+|S_2|^2 \right) = \frac{1}{4\pi}
\frac{1-g^2}{\left(1+g^2 -2 g \cos\theta \right)^{3/2}},
\label{hgphasefunc}
\end{equation}
with $ -1 < g < 1$, and
\begin{equation}
\frac{\left(|S_2|^2-|S_1|^2 \right)}{\left(|S_1|^2+|S_2|^2 \right)} = 
-p_{max}\frac{\sin^2\theta}{1+\cos^2\theta},
\label{rayleighpol}
\end{equation}
where for consistency with Eq. (\ref{sandptostokes}) the PA of the
electric vector is measured from the scattering plane.  For Rayleigh
scattering $0 \le p_{max} \le 1$, and $Q < 0$.  In the case of Mie
scattering from dielectric spheres, Eqs. (\ref{hgphasefunc}) and
(\ref{rayleighpol}) are a satisfactory approximation for grains with
$x \lesssim 1$.  This recipe cannot describe a second peak at $\theta
= \pi$ associated with enhanced backscatter or give polarization
parallel to the scattering plane. Nonetheless, it has several
desirable features: 1) there are only two adjustable grain
parameters---the other model parameters are the normalization, the
inner and outer disk radii, and the power law slope of the radial dust
distribution; 2) the computational simplicity of evaluating
Eqs. \ref{hgphasefunc} and \ref{rayleighpol}.  This is a consideration
as performing simultaneous, non-linear least-squares fits requires
multiple ($\sim 10^5$) numerical evaluations of the integral in
Eq. (\ref{sb_integral}).

Figure \ref{hggzodi} shows the least-squares fit to the
Henyey-Greenstein model and Table \ref{model_table} lists the fit
parameters and goodness-of-fit, $\chi^2_\nu$.  The fit was found using
Craig Markwardt's constrained, non-linear least squares program,
{\tt MPFIT}\footnote{\url{http://cow.physics.wisc.edu/$\sim$craigm/idl/}}, 
which
is implemented in the 
IDL programming language\footnote{\url{http://www.ittvis.com}}.  
As we are fitting both the surface brightness and the polarization,
the results of the fit depend on the relative weights attributed to
each data set.  For the current observations the errors in the surface
brightness are smaller than for the fractional polarization.
Therefore, we have assigned a minimum 10\% fractional uncertainty to
the surface brightness to reflect our prejudice that the polarization
data carry significant information regarding the nature of the grains.
This assumption is justified by noting that NE and SW wings of the
disk are not identical---these local variations in the dust column,
e.g., caused by density waves or the injection of fresh material,
cannot be fit by our simple power law model, and therefore these
deviations should not contribute to $\chi^2_\nu$.

The fit achieves $\chi^2_\nu = 1.7$ with $\nu = 146$ degrees of
freedom, which is gratifyingly good given that there are only six
parameters.  The model accurately reproduces the shape of the surface
brightness profile, the steep rise in polarization between 20 and 40
AU, and is consistent, within the errors, with the leveling off at $p
\simeq 0.40$ beyond this point.  Key aspects of the fit include highly
polarizing grains, $p_{max} = 0.53\pm0.03$, strong forward scattering,
$g = 0.68\pm 0.01$ and an inner hole at $r_1 = 38\pm0.5$ AU.  The
quoted errors are only the formal errors and should be treated with
some caution.  For example, the best-fit value of $r_2$ is biased by
the fact that our last data point lies at 80 AU. If the outer radius
is held fixed at 200 AU, then $\chi^2_\nu$ increases to 1.76, which is
unacceptable only at the 1-$\sigma$ level and perhaps indicative that
a single grain population cannot account for the scattered light from
the entire disk. Back scattering, which is typical of particles with
$x \gtrsim 1$ can be described by a simple extension of
Eq. (\ref{hgphasefunc}) to a two component Henyey-Greenstein function
$H(\theta) = (1-B) H(\theta,g_1) + B H(\theta,g_2)$, where $0 \le g_1
< 1$ and $-1 < g_2 \le 0$.  The introduction of two extra scattering
parameters does not achieve an improved fit, and therefore we find no
evidence for enhanced backscatter.

These disk models teach us that the radial dust distribution and the
phase function are covariant if only Stokes $I$ is available.
Evidently, a uniform disk with a high degree of forward scattering can
mimic a disk with a steep decline in grain opacity that is combined
with more isotropic scattering.  The results of analyses that adopt a
specific radial profile, e.g., by fixing $\beta$, must be interpreted
accordingly \citep[e.g.,][on $\beta$ Pic.]{2006AJ....131.3109G}.  For
parameterized grain properties, e.g., Eqs. (\ref{hgphasefunc}) and
(\ref{rayleighpol}) this covariance remains.  When a physical
scattering model is adopted, which ties together $g$ and $p_{max}$,
and $I$ and $Q$ are fitted simultaneously, then this degeneracy is
broken.

\subsection{A Zodiacal Dust Model}

Now that we have shown that a simple model can reproduce the
observations of the AU Mic disk we can ask whether grains with the
optical scattering properties consistent with experimental studies of
Solar System dust work too.  We adopt \citet{1985A&A...146...67H}'s
description of the scattering phase function inferred from the
observed angular variation of the surface brightness and polarization
of the Zodiacal light. Hong's formulation is convenient because it
represents the scattering characteristics of interplanetary particles
as a three component linear combination of three Henyey-Greenstein
functions.

\begin{figure}
\plotone{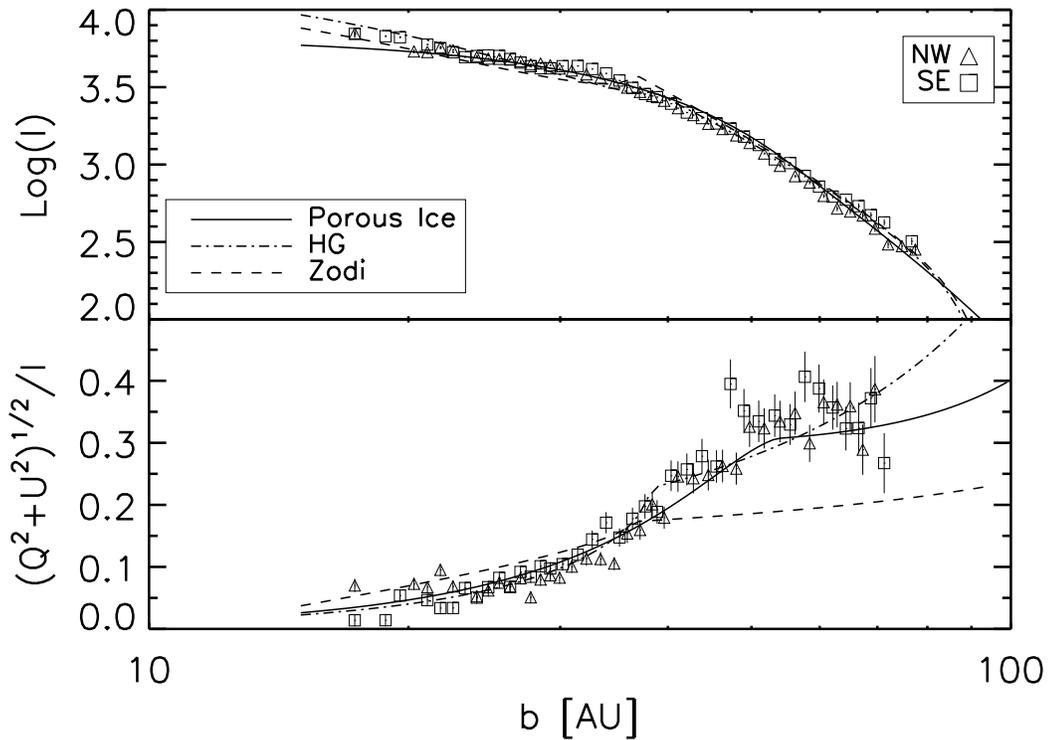}
\caption{Simultaneous fits to the surface brightness profile (top) and
the degree of linear polarization (bottom).  Three different models
are shown (see Table 1 for details).  The grains in the best fit model
(solid line) are porous (91\%) water ice with $\chi^2_\nu = 1.6$.  The
dash-dot line is a single-component Henyey-Greenstein model
$\chi^2_\nu = 1.7$.  The dashed line is a Zodiacal dust model.  The
Zodiacal dust cannot account for the observed polarization fraction
($\chi^2_\nu = 6.3$).  The phenomenological Henyey-Greenstein and the
physical porous models can explain both observations. They have in
common a high degree of forward scattering and polarization.  Models
where the dust grains are solid yield unacceptable fits.  The error
bars in the lower panel include the systematic uncertainty in the
degree of polarization.  }
\label{hggzodi}
\end{figure}

This example reproduces the calculation that
\citet{1997AREPS..25..175A} used to describe the polarization
signature of $\beta$ Pic.  There are now only four free parameters:
the normalization of the vertical optical depth, the radii of the
inner and outer holes and the power law index of the radial density
distribution.  Figure \ref{hggzodi} shows that this model fits the
surface brightness profile well, but fails to provide an adequate
description of the measured polarization. Chi-squared for the combined
data set is 6.29, which can be rejected with high confidence
($>$99\%).  The Zodiacal dust model cannot explain the steep rise in
polarization over the inner disk (20--50 AU) and it cannot account for
the high polarization in the outer disk.  The radial extent of the
disk is similar to that of the Henyey-Greenstein model with an inner
hole at $37\pm1$ AU. The disk terminates at $r_2$ = 90 AU.  Since the
scattering asymmetry parameter is fixed at $g = 0.4$ the radial
opacity distribution is steeper than in the Henyey-Greenstein model. A
satisfactory fit to the polarization data cannot be found, even if we
set the weights for the surface brightness to zero.

The lessons from these results are twofold. First, surface brightness
data alone are insufficient to constrain grain optical properties and
their radial distribution.  Second, particles with the optical
properties of Zodiacal grains cannot explain the polarization
signature of AU Mic. Thus, the conclusions of the qualitative
discussion in \S \ref{onedprofiles} are not fully borne out: simply
changing the radial dust distribution does not explain the difference
between $\beta$ Pic and AU Mic.

\clearpage
\onecolumngrid

\twocolumngrid

Choosing an alternate type of Solar System dust, e.g., cometary
grains, does not significantly improve the discrepancy between the
model and the polarization data because cometary dust does not have
sufficiently high peak linear polarization.  The dusty comets, of
which Comet 1996 B2 (Hyakutake) is typical, show little dispersion in
their polarizing properties.  For example, the maximum degree of
linear polarization of Comet 1996 B2 (Hyakutake) was observed to be
0.24 and 0.26 at 484 and 684 nm, respectively, at a phase angle near
94 degrees \citep{1998Icar..133..286K}.

\subsection{Dielectric Spheres: Mie Theory}
\label{mie_theory}

Several debris disk studies have used Mie theory to evaluate the
complex elements of the amplitude scattering matrix
\citep[e.g.,][]{1999A&A...352..508V, 2000ApJ...539..424K}.  Solar
System dust particles are not spheres, and computing the cross
sections using Mie theory may be misleading
\citep[cf.][]{1994AREPS..22..553G}.  However, Zodiacal and cometary
dust have too low a value of peak linear polarization to be consistent
with the AU Mic data.  As polarization efficiency increases with
decreasing particle size, with Rayleigh scatterers representing the
limiting case, it is worth investigating whether small dielectric
spheres composed of common astrophysical material can be invoked.
Perhaps erosion in AU Mic's disk is so severe that the grains have
been ground down to their constituent interstellar precursors?  Since
internal grain structure can be neglected as $x \rightarrow 0$ the Mie
approximation should be sufficiently accurate to explore this
possibility.

Mie models with small grains can explain the observed polarization.
Adopting astronomical ``silicate'' ($m = 1.65-0.01i $) as the grain
material yields a joint fit that is better than the Zodiacal dust
model, but the best value of $\chi^2_\nu = 4.4$, is clearly
unacceptable at a high level of confidence ($>$ 99\%).  Other
parameters are listed in Table \ref{model_table}. Most of the
contributions to $\chi^2_\nu$ are from residuals relative to the
surface brightness profile, which are attributable to grain scattering
that is too isotropic.  Only a narrow range of spherical particle
sizes come close to approximating the data because the joint fit
simultaneously constrains the phase function and the maximum
polarization---quantities that vary rapidly with particle size.  The
best fit size parameter is $x = 1.63 \pm0.01$ for astronomical
``silicate'', or $a = 0.16\; \mu$m.  Adopting dirty water-ice grains
($m=1.33-0.01i$) reduces $\chi^2_\nu$ significantly, but not to an
acceptable level.  Organic refractory material
\citep[$m=1.98-0.28i$;][]{1997A&A...323..566L} fares worse than either
rock or ice.

\begin{table}[th]
\caption{Disk Models}
\begin{center}
\begin{tabular}{lcccccc}
\hline
Model   & $\beta$ & $r_1$  & $r_2$ & $p_{max}$  & $g$ &   $\chi^2_\nu$  \\
\hline
\hline
Porous water ice$^a$  &  3.02   &   53.1  &  177.8  &  0.94     &  0.81   & 1.6 \\
                 & (0.5)$^b$   &   (2.1) &  (69.7) &     &   & \\
HG$^c$  &  0.90   &   38.0  &   92.5  &  0.53     &  0.68   &  1.7 \\
         & (0.25)  &   (0.5) &   (4.7) &  (0.02)   &(0.01)   &           \\

HG$^d$  &  2.47   &   41.6  &  200    &  0.62     &  0.71   &  1.8   \\
         & (0.13)  &  (0.4)  & $\dots$ & (0.03)    &  (0.01)               \\
Water ice$^e$  &  2.14    &  41.6   &   100.4  &  0.47  & 0.68  & 3.1  \\
          & (0.26)   &  (0.8)  &   (3.3) &   &  &       \\
ISM$^f$   &  1.79    &  35.8   &  87.1   &  0.42  & 0.45     & 4.0 \\
          & (0.27)   & (0.4)   & (4.4)   & $\dots$    & $\dots$           \\
Silicate$^g$ & 1.37  &  34.1   &   85.9  &  0.38      & 0.59  & 4.4  \\
          & (0.26)   &  (0.5)  &   (3.3) &  (0.04)    & (0.01)  &     \\
SS Zodi$^h$ & 1.68   &   36.6  & 94.3    &   0.25   &  0.40     &   6.3   \\
            & (0.49) &   (0.8) & (13.8)  &  $\dots$ & $\dots$   &        \\
\hline
\end{tabular}
\end{center}
(a) Maxwell-Garnett/Mie model for porous water ice with $m = 1.33-0.01i$. 
    The best fit porosity is $0.91\pm0.09$.
    The best fit grain size corresponds to $x = 3.26 \pm 0.15$ ($620 \pm 30$~nm diameter)
    The peak linear polarization $p_{max}$ asymmetry parameter $g$ are 
    derived parameters. \\
(b)  Formal 1-$\sigma$ uncertainties are indicated in 
parentheses. Derived parameters that are not 
model parameters have a blank in the second row. 
The use of ``$\dots$''  implies
that the corresponding parameter is fixed. \\
(c) Single component Henyey-Greenstein model. \\
(d) Single component Henyey-Greenstein model with $r_2$ fixed. \\
(e) Single particle Mie model with best fit size parameter, $x=2.13\pm 0.01$ 
for solid ``dirty ice'' grains ($m = 1.33 +0.01i$). \\
(f) Interstellar dust model (White 1979). \\
(g) Single particle Mie model with best fit size parameter, $x=1.63\pm 0.01$ 
for solid ``silicate'' grains ($m = 1.65 -0.01i$). \\
(h) Solar System Zodiacal three-component Henyey-Greenstein dust model (Hong 1985). \\
\label{model_table}
\end{table}

The conclusion that $\chi^2_\nu$ varies significantly with choice of
$n$, the real part of the refractive index, suggests that it should be
adopted as a fit parameter. Such models have six free parameters---the
same as the Henyey-Greenstein model of \S \ref{hgmodel}.  A
satisfactory fit, with $\chi^2_\nu=1.6$, is achieved for $n = 1.03\pm
0.03$ and $x = 3.25\pm 0.18$ (solid line in Fig. \ref{hggzodi}).
Based on the value of $\chi^2_\nu$, this low-index Mie fit is slightly
better than the Henyey-Greenstein model, and this model is consistent
with the disk extending beyond 200 AU. While our choice of the complex
part of the refractive index is somewhat arbitrary, making the grains
more or less absorbing does not qualitatively change our conclusions.

Using our best fit model we also investigate whether or not the inner
disk ($r < r_1$) is dust free.  By adding a parameter that describes
an inner hole with constant vertical depth, we find that $\tau_\perp
(r < r_1) < 0.003\;\tau_\perp(r_1)$ (99\% confidence).  The inner disk
is devoid of micron-sized grains, which according to the
\citet{2005astro.ph.10527S} model means that collisions dominate,
i.e., this is a ``Type B'' disk where grains that are dragged inward
by corpuscular and Poynting-Robertson drag undergo a destructive
collision.

\section{Discussion}
\label{discussion}

\begin{figure}[t]
\plottwo{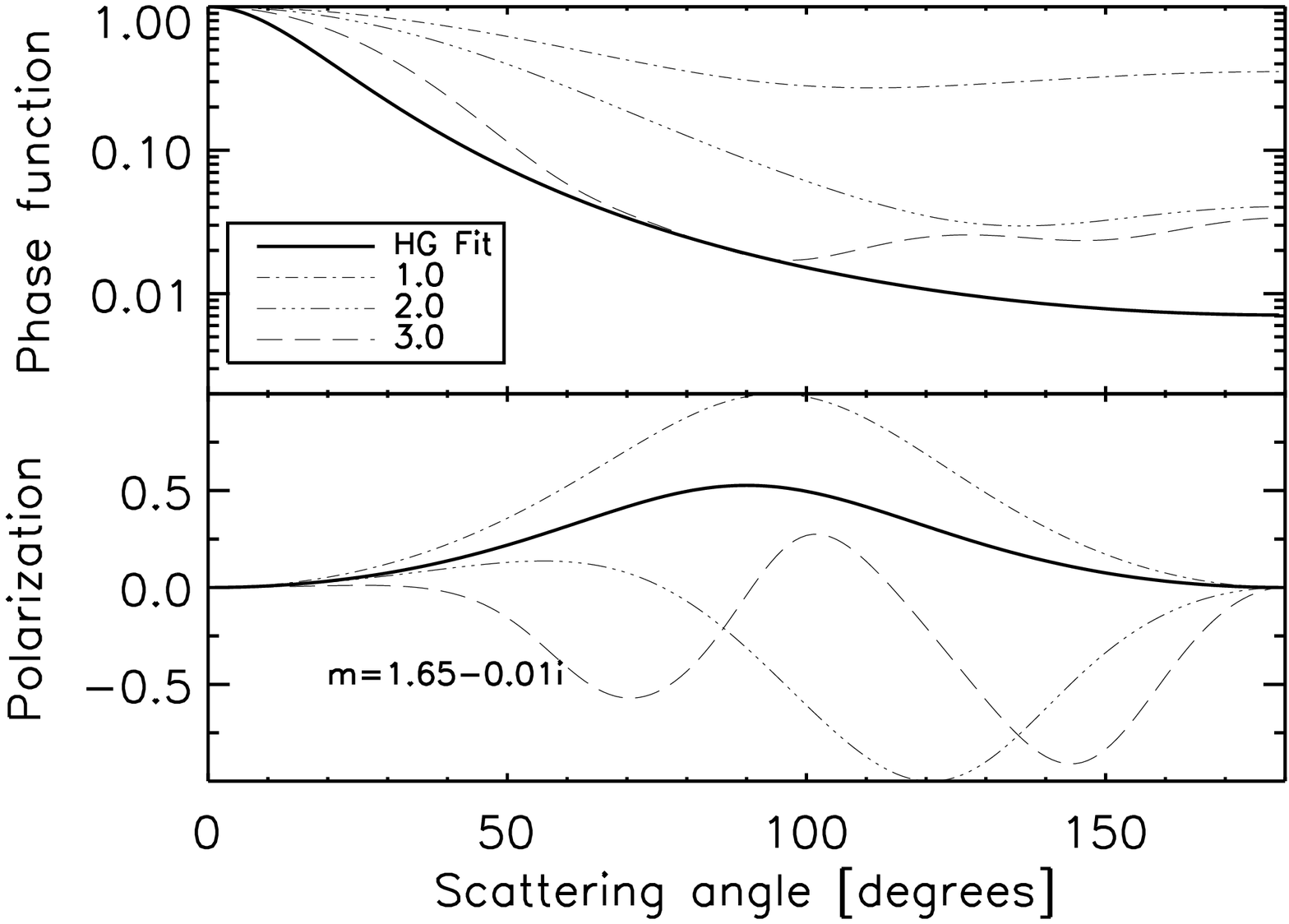}{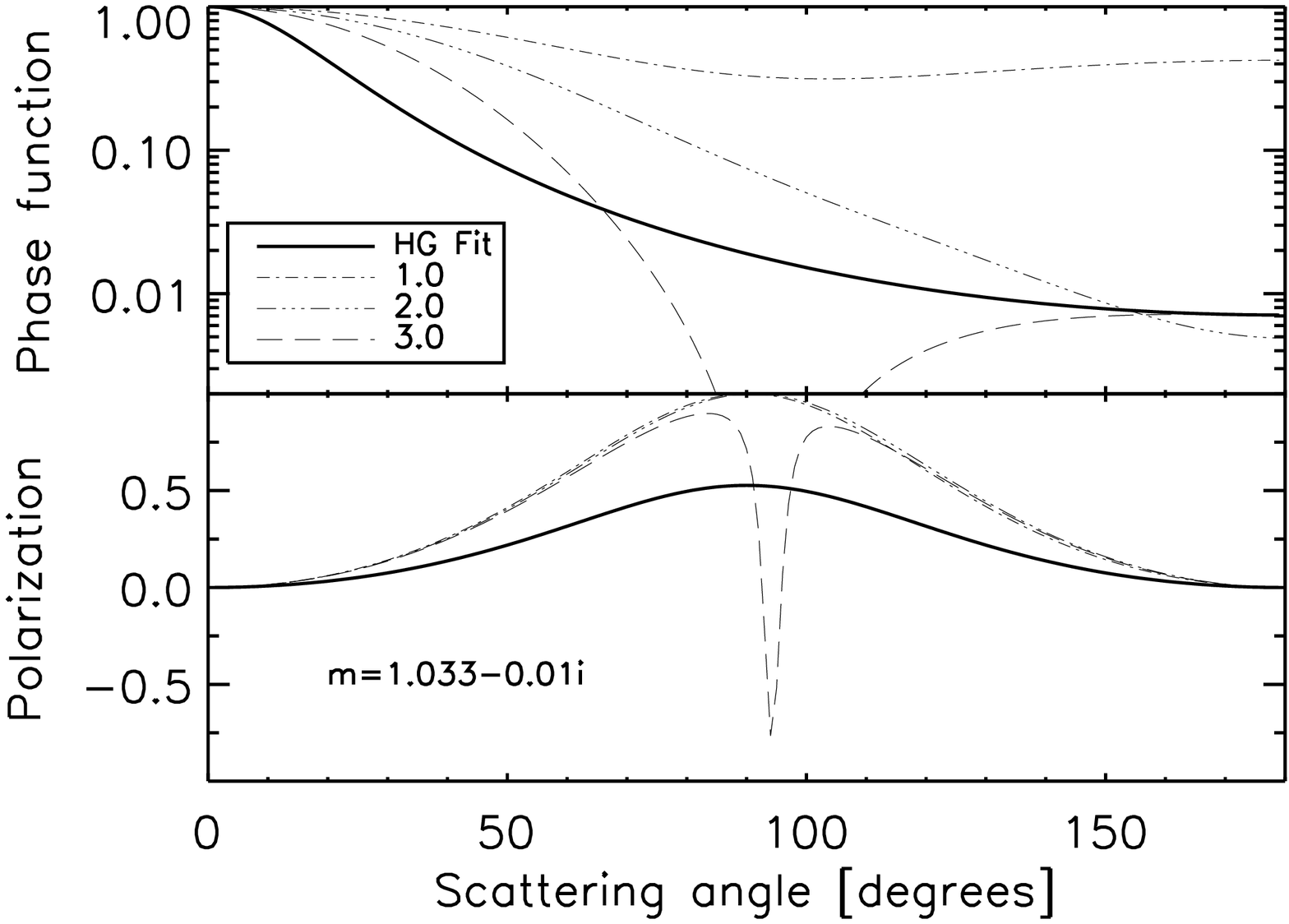}
\caption{Normalized phase function (top) and polarization fraction
(bottom) versus scattering angle. The heavy solid line is inferred
from the best fit single component Henyey-Greenstein model with
scattering asymmetry parameter $g=0.68$.  Here polarization denotes
$-Q/I$, thus negative polarization indicates that the electric field
is oriented parallel to the scattering plane.  On the left are the
results of Mie calculations for spheres with $x$ = 1, 2 and 3 with
conventional $m = 1.65-0.01i$ (``silicate''). The best fit single
particle size Mie fit (Table 1), has $x = 1.63\pm 0.01$.  The
inability of a sphere to simultaneously match the phase function and
the polarization explains why this model can only achieve a poor fit.
Small dielectric spheres with $x \simeq 3$ can account for the strong
forward scattering, but they cannot simultaneously provide a high
degree of polarized light perpendicular to the scattering plane.  On
the right is shown the result when the real part of the refractive
index is allowed to be a free parameter.  Grains with $m =
1.033-0.01i$ and $x = 3.25$ provide a satisfactory fit to the
data---at least as good as the one component Henyey-Greenstein
model. The low index means that the grains must be extremely porous
($\ge 90$\%). }
\label{phase_function}
\end{figure}

Figure \ref{hggzodi} and the corresponding fit parameters in Table
\ref{model_table} demonstrate that a variety of radial grain
distributions can account for the observed surface brightness,
although a large inner hole with radius of 40--50 AU is common to all
models.  Taken together, the surface brightness and degree of linear
polarization narrow down the range of acceptable grain optical
properties.  Inspection of Figure \ref{phase_function} shows that a
combination of strong forward scattering and a high polarizing
efficiency, with a Rayleigh-like $\sin^2\theta/(1+\cos^2\theta)$
angular variation is sufficient to describe these data.  Our analysis
implies with high confidence that the constituent dust grains exhibit
high maximum linear polarization ($p_{max} \ge 0.50$) and strong
forward scattering ($g \ge 0.7$).

Plots of the phase function and polarization fraction versus phase
angle shown in Figure \ref{phase_function} help highlight the common
features of the statistically acceptable Henyey-Greenstein and
low-index Mie models, and how their scattering properties differ from
conventional rocky or ice-grain models. Large rocky or icy grains
typically have strong forward scattering, but polarization which
oscillates with phase angle. Therefore, $Q$ tends to average to zero
along a line of sight that integrates over a range of scattering
angles (refer to \S \ref{linpol}).  As the conventional Mie fit shows,
small grains are good polarizers, but scatter too isotropically to be
consistent with the data.  The low index Mie model suggests a physical
scenario that combines strong forward scattering and high
polarization.

Application of the Clausius-Mossotti relation
\citep{1962clel.book.....J} implies that such a low value of $n$ must
be associated with a very porous medium---the only terrestrial analog
that comes to mind is silica aerogel.  Aerogels are transparent,
highly porous materials of low density, ranging from 0.05 to 0.15
g~cm$^{-3}$, with a corresponding refractive index of 1.01 to 1.04,
respectively.  Aerogel has porosity on a micron scale and is composed
of individual silica grains with diameters of $\simeq$ 10 nm, which
are linked in a highly porous dendritic backbone.  Although aerogel is
produced in a process that is unlikely to occur in an astrophysical
setting---hydrolysis of methyl silicate in the presence of a solvent
(ethanol) that is subsequently evaporated at high temperature and
pressure---the comparison is not entirely frivolous.  The appearance
of aerogel is often characterized as ``solid blue smoke'', because to
a good approximation the scattering is Rayleigh scattering
\citep{1993JQWIISpectroscRadiarTransfer50}.  Thus, aerogel is an
example of a bulk material that interacts with electromagnetic
radiation in a way that is determined by its microscopic structure.

Allowing the real part of the refractive index to vary as described in
\S \ref{mie_theory} while holding the imaginary part fixed violates
the Kramers-Kronig relation.  An approach that has a better physical
basis is to use an effective medium theory to compute the optical
behavior of a porous particle described as vacuum matrix ($n=1$) with
embedded inclusions \citep{2003pid..book.....K}.  Using the
Maxwell-Garnett rule we can choose a refractive index for the bulk
material and use the grain porosity as a model parameter.  The
best-fit porous grain model (see Fig. \ref{hggzodi} and the first line
of Table \ref{model_table}) is practically identical to the variable
index fit, yielding essentially the same structural parameters.  The
grain porosity is 91--94\%, depending on whether we assume that the
matrix from which our grains are made is ice or rock.

\clearpage

\onecolumngrid

\twocolumngrid

As our aerogel analogy reminds us, porous materials are likely highly
anisotropic, and we may not be free to assume that we can neglect the
effects of nearest neighbors within the matrix.  We therefore examine
the results of numerical calculations (e.g., discrete-dipole
approximation and transition-matrix) of light scattering by aggregates
to understand whether or not our interpretation of the Mie results in
terms of porous grains is credible.

A lucid exposition of the transition-matrix method applied to
composite interstellar grains is given by \citet{2004ApJ...615..286I}.
Using this method \citet {2000Icar..148..526P} show results for two
instances of silicate ($m = 1.65 -0.01i$) grain clusters consisting of
31 particles or ``monomers'' each with $x_m = 1.5$.  A relatively
compact aggregate with approximately 70\% porosity has $g = 0.75$ and
$p_{max} = 0.52$.  The more porous particle (81\%) has $g = 0.75$ and
$p_{max} = 0.65$.  In neither case does the degree of polarization
oscillate with phase angle.  These clusters have optical properties
which make them promising analogs of the material inferred to dominate
the AU Mic disk.

\begin{figure}
\plotone{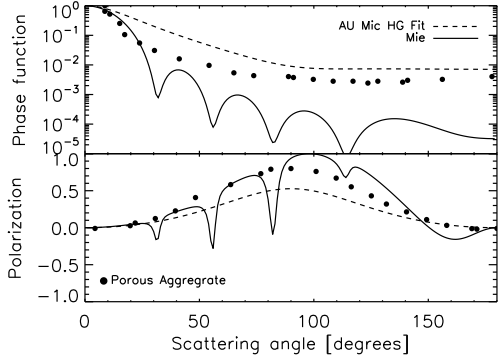}
\caption{The phase function and polarization (dots) for a large ($x =
  10.2$) porous (90\%) silicate particle ($m=1.6-0.01i$) composed of
  128 small ($x=0.9$) monomers from \citet{2006A&A...449.1243K}.  This
  particle has $g = 0.84$ and $p_{max} = 0.82$.  The dashed line is
  the phase function ($g=0.68$) and polarization ($p_{max} = 0.53$)
  for the Henyey-Greenstein model which best fits the AU Mic disk. 
  The porous grain is more
  forward scattering and more polarizing than required by the AU Mic
  data, but nonetheless its optical properties show that highly porous
  aggregates 
  constitute an excellent candidate for the AU Mic
  disk.  Also shown is the Mie approximation for the porous grain with
  dielectric properties derived using the Clausius-Mossotti relation
  ($m = 1.047-0.0007i$).  The Mie calculation is a useful 
  first approximation to the polarization, but overestimates $g$ by 
  about 15\%.  }
\label{aggregate_vs_mie}
\end{figure}

\citet{2006A&A...449.1243K} present additional
results for larger, more porous aggregates.  Figure
\ref{aggregate_vs_mie} shows the phase function and polarization for a
large ($x_c = 10.2$) porous (90\%) silicate cluster ($m=1.6-0.01i$)
composed of 128 small ($x_m=0.9$) monomers.  This particle has $g =
0.84$ and $p_{max} = 0.82$, and has optical properties which make it
an excellent candidate material for the AU Mic disk.  Also shown is
the corresponding Mie calculation with dielectric properties derived
using the Clausius-Mossotti relation.  It is evident that the Mie
calculation is only a rough approximation---the polarization curve is
reasonably well reproduced, and $g$ is overestimated by about 15\%.
Although the qualitative conclusion that implicates porous grains is
secure it seems unlikely that the accuracy of effective medium
theories is sufficient, for example, to distinguish between different
coagulation schemes that are characterized by different porosity.  It
will be necessary to abandon Mie theory in favor of numerical modeling
of aggregate scattering in the next stage of debris disk modeling.

If highly porous aggregate grains explain the polarization signature
of AU Mic's debris disk, and if the dust beyond $r_1\simeq 40$ AU
originates from a ``birth ring'' of parent bodies $\lesssim $10 cm in
size, as envisioned by \citet{2005astro.ph.10527S}, then this porosity
may be a signature of the agglomeration process whereby interstellar
grains first grew into macroscopic sized objects.  In the inner Solar
System porous particles occur naturally in cometary dust, where the
sublimation of ices leaves a ``bird's nest'' of refractory organic and
silicate material \citep{1990ApJ...361..260G}. Porous grains in the
$\beta$ Pic disk may originate from cometary activity
\citep{1997A&A...323..566L}. However, the birth ring in AU Mic lies
safely outside the $\sim$1 AU ice sublimation point.  Based on
collisional lifetime arguments, the size of the parent-bodies that
supply the observed dust in AU Mic is in the decimeter range
\citep{2005astro.ph.10527S}.  Though the existence of larger bodies
that will suffer compaction and restructuring
\citep{2000Icar..143..138B} is not excluded, they are not the dominant
reservoir for dust observed at optical and near-infrared wavelengths.
Evidently, shock compression during attrition of the parent bodies in
the birth ring is not significant. We envision these bodies as so
weakly bound that even the most glancing collisions lead to their
disruption.  Recent laboratory studies of particle coagulation in the
proto-solar nebula by ballistic cluster-cluster aggregation
\citep{1998Icar..132..125W} leads to the formation of highly ($>$90\%)
porous aggregates.  Our evidence suggests that such a process may have
mediated the initial growth of planetesimals.

Observations of scattered light at a single wavelength are primarily
sensitive to grains with $x \simeq 1$, and do not place strong
constraints on the particle size distribution.  However, preliminary
calculations show that the measured polarization is consistent with a
Dohnanyi spectrum (\citet{fitz06} and L. E. Strubbe 2006, private
communication).  Polarization measurements in the UV through the
near-infrared could be used to measure grain porosity as a function of
grain size.

\section{Summary}
\label{summary}

We have observed the AU Mic debris disk at F606W (broad $V$) with the
POLV polarizing filter set in ACS high resolution camera aboard HST.
The coronagraph and PSF subtraction were used to suppress scattered
light.  The disk light is polarized, with the degree of linear
polarization rising steeply from 0.05 to 0.35 between 20 and 50 AU and
reaching a maximum of about 0.40 within 80 AU.  The inner and outer
working limits are set by systematic errors in PSF subtraction and
declining SNR, respectively.  The linear polarization is oriented with
the electric field perpendicular to the disk, which is characteristic
of scattering by optically thin, small grains.

We have factored systematic errors in the ACS polarization
measurements into our uncertainties.  For the bright, strongly
polarized emission between 35 and 55 AU these errors and not
measurement errors dominate.  However, the ACS/HRC polarization
calibration campaign is on-going, and analysis of the resultant data,
including a full Mueller matrix description for the HRC+F606W+POLV
combination (cf. the \citet{2000PASP..112..983H} analysis of NICMOS)
and application to this data set will improve the reliability and
fidelity of these results.

AU Mic and $\beta$ Pic have different polarization signatures.
$\beta$ Pic shows a shallower gradient and lower peak in polarization
fraction.  We attribute this different to two factors: 1) the two
disks are probed on different spatial scales relative to their inner
and outer boundaries; 2) the grains in the AU Mic disk have a higher
peak linear polarization than those of $\beta$ Pic.  We place limits
on the radial distribution of grains and their optical scattering
properties by performing simultaneous fits to the observed surface
brightness and the degree of polarization.  These fits show that the
inner boundary of the AU Mic disk is located between 40--50 AU, and
the dust component which is responsible for the strong linear
polarization extends to 100--150 AU.  The uncertainty occurs primarily
because the inferred spatial structure of the disk and the grain
optical properties---the scattering asymmetry factor $g$ and the peak
linear polarization $p_{max}$---are covariant if $g$ and $p_{max}$ are
independent.  We can state with good confidence that $g\ge 0.7$ and
$p_{max}> 0.50$. The inner disk is virtually free of micron-sized
grain, and Type B conditions prevail
\citep[cf.][]{2005astro.ph.10527S}.

This combination of optical properties occurs naturally in porous
media.  Once we adopt a physical description for the electromagnetic
properties of the scatterers, $g$ and $p_{max}$ are not independent
and the covariance with the radial dust distribution is greatly
reduced.  Our best fit physical model, which invokes Mie theory and
low index grains, implies that the inner regions of the AU Mic disk
($<$ 50 AU) are depleted of small grains.  This finding lends support
to the \citet{2005astro.ph.10527S} birth ring.

The best fit, porous grain model using Mie theory and the
Maxwell-Garnett rule implies a grain porosity of 91--94\%, depending on
whether the grain matrix is ice or rock.  Porous grains are a natural
consequence of particle growth. However, the accuracy of the effective
medium theory, which we used to convert the dielectric constant into a
porosity, is probably insufficient to favor one growth mechanism over
another, e.g., cluster-cluster vs. cluster-particle agglomeration.
Better approximations for calculating the optical properties of
clusters, e.g., the discrete-dipole approximation or the transition
matrix method must be employed.

\begin{acknowledgements}

We thank John Biretta for help with understanding the polarization
calibration ACS. This work is based on observations with the NASA/ESA
Hubble Space Telescope obtained at the Space Telescope Science
Institute (STScI), which is operated by the Association of
Universities for Research in Astronomy.  Support for Proposal number
GO-10228 was provided by NASA through a grant from STScI under NASA
contract NAS5-26555.  This work has been supported by the National
Science Foundation Science and Technology Center for Adaptive Optics,
managed by the University of California at Santa Cruz under
cooperative agreement No. AST-9876783. B.C.M. acknowledges
support from the National Research Council of Canada. 

\end{acknowledgements}


\begin{thebibliography}{50}
\expandafter\ifx\csname natexlab\endcsname\relax\def\natexlab#1{#1}\fi

\bibitem[{{Artymowicz}(1997)}]{1997AREPS..25..175A}
{Artymowicz}, P. 1997, Annual Review of Earth and Planetary Sciences, 25, 175

\bibitem[{{Augereau} \& {Beust}(2006)}]{2006astro.ph..4313A}
{Augereau}, J.~C. \& {Beust}, H. 2006, ArXiv Astrophysics e-prints

\bibitem[{{Aumann} {et~al.}(1984){Aumann}, {Beichman}, {Gillett}, {de Jong},
  {Houck}, {Low}, {Neugebauer}, {Walker}, \& {Wesselius}}]{1984ApJ...278L..23A}
{Aumann}, H.~H., {Beichman}, C.~A., {Gillett}, F.~C., {de Jong}, T., {Houck},
  J.~R., {Low}, F.~J., {Neugebauer}, G., {Walker}, R.~G., \& {Wesselius}, P.~R.
  1984, \apjl, 278, L23

\bibitem[{{Backman} \& {Paresce}(1993)}]{1993prpl.conf.1253B}
{Backman}, D.~E. \& {Paresce}, F. 1993, in Protostars and Planets III, ed.
  E.~H. {Levy} \& J.~I. {Lunine}, 1253--1304

\bibitem[{{Barrado y Navascu{\'e}s} {et~al.}(1999){Barrado y Navascu{\'e}s},
  {Stauffer}, {Song}, \& {Caillault}}]{1999ApJ...520L.123B}
{Barrado y Navascu{\'e}s}, D., {Stauffer}, J.~R., {Song}, I., \& {Caillault},
  J.-P. 1999, \apjl, 520, L123

\bibitem[{{Biretta} \& {Kozhurina-Platais}(2004)}]{2004biretta-10}
{Biretta}, J. \& {Kozhurina-Platais}, V. 2004, {ACS Polarization
  Calibration---II. The POLV Filter Angles: ACS 2004-10} (Baltimore: Space
  Telescope Science Institute)

\bibitem[{{Biretta} {et~al.}(2004){Biretta}, {Kozhurina-Platais}, {Boffi},
  {Sparks}, \& {Walsh}}]{2004biretta-09}
{Biretta}, J., {Kozhurina-Platais}, V., {Boffi}, F., {Sparks}, W., \& {Walsh},
  J. 2004, {ACS Polarization Calibration---I. Introduction and Status Report:
  ACS 2004-09} (Baltimore: Space Telescope Science Institute)

\bibitem[{{Blum} \& {Wurm}(2000)}]{2000Icar..143..138B}
{Blum}, J. \& {Wurm}, G. 2000, Icarus, 143, 138

\bibitem[{{Chandrasekhar}(1960)}]{1960ratr.book.....C}
{Chandrasekhar}, S. 1960, {Radiative transfer} (New York: Dover, 1960)

\bibitem[{{Chen} {et~al.}(2005){Chen}, {Patten}, {Werner}, {Dowell},
  {Stapelfeldt}, {Song}, {Stauffer}, {Blaylock}, {Gordon}, \&
  {Krause}}]{2005ApJ...634.1372C}
{Chen}, C.~H., {Patten}, B.~M., {Werner}, M.~W., {Dowell}, C.~D.,
  {Stapelfeldt}, K.~R., {Song}, I., {Stauffer}, J.~R., {Blaylock}, M.,
  {Gordon}, K.~D., \& {Krause}, V. 2005, \apj, 634, 1372

\bibitem[{{Evans} {et~al.}(1957){Evans}, {Menzies}, \&
  {Stoy}}]{1957MNRAS.117..534E}
{Evans}, D.~S., {Menzies}, A., \& {Stoy}, R.~H. 1957, \mnras, 117, 534

\bibitem[{{Fitzgerald} {et~al.}(2006){Fitzgerald}, {Kalas}, {Duch{\^e}ne},
  {Pinte}, \& {Graham}}]{fitz06}
{Fitzgerald}, M.~P., {Kalas}, P.~G., {Duch{\^e}ne}, G., {Pinte}, C., \&
  {Graham}, J.~R. 2006, to be submitted to \apj, 000, 000

\bibitem[{{Ford} {et~al.}(2003){Ford}, {Clampin}, {Hartig},
  {et~al.}}]{2003SPIE.4854...81F}
{Ford}, H.~C., {Clampin}, M., {Hartig}, G.~F., {et~al.} 2003, in Future EUV/UV
  and Visible Space Astrophysics Missions and Instrumentation. Edited by J.
  Chris Blades, Oswald H. W. Siegmund. Proceedings of the SPIE, Volume 4854,
  pp. 81-94 (2003)., ed. J.~C. {Blades} \& O.~H.~W. {Siegmund}, 81--94

\bibitem[{{Gledhill} {et~al.}(1991){Gledhill}, {Scarrott}, \&
  {Wolstencroft}}]{1991MNRAS.252P..50G}
{Gledhill}, T.~M., {Scarrott}, S.~M., \& {Wolstencroft}, R.~D. 1991, \mnras,
  252, 50P

\bibitem[{{Golimowski} {et~al.}(2006){Golimowski}, {Ardila}, {Krist},
  {Clampin}, {Ford}, {Illingworth}, {Bartko}, {Ben{\'{\i}}tez}, {Blakeslee},
  {Bouwens}, {Bradley}, {Broadhurst}, {Brown}, {Burrows}, {Cheng}, {Cross},
  {Demarco}, {Feldman}, {Franx}, {Goto}, {Gronwall}, {Hartig}, {Holden},
  {Homeier}, {Infante}, {Jee}, {Kimble}, {Lesser}, {Martel}, {Mei},
  {Menanteau}, {Meurer}, {Miley}, {Motta}, {Postman}, {Rosati}, {Sirianni},
  {Sparks}, {Tran}, {Tsvetanov}, {White}, {Zheng}, \&
  {Zirm}}]{2006AJ....131.3109G}
{Golimowski}, D.~A., {Ardila}, D.~R., {Krist}, J.~E., {Clampin}, M., {Ford},
  H.~C., {Illingworth}, G.~D., {Bartko}, F., {Ben{\'{\i}}tez}, N., {Blakeslee},
  J.~P., {Bouwens}, R.~J., {Bradley}, L.~D., {Broadhurst}, T.~J., {Brown},
  R.~A., {Burrows}, C.~J., {Cheng}, E.~S., {Cross}, N.~J.~G., {Demarco}, R.,
  {Feldman}, P.~D., {Franx}, M., {Goto}, T., {Gronwall}, C., {Hartig}, G.~F.,
  {Holden}, B.~P., {Homeier}, N.~L., {Infante}, L., {Jee}, M.~J., {Kimble},
  R.~A., {Lesser}, M.~P., {Martel}, A.~R., {Mei}, S., {Menanteau}, F.,
  {Meurer}, G.~R., {Miley}, G.~K., {Motta}, V., {Postman}, M., {Rosati}, P.,
  {Sirianni}, M., {Sparks}, W.~B., {Tran}, H.~D., {Tsvetanov}, Z.~I., {White},
  R.~L., {Zheng}, W., \& {Zirm}, A.~W. 2006, \aj, 131, 3109

\bibitem[{{Greenberg} \& {Hage}(1990)}]{1990ApJ...361..260G}
{Greenberg}, J.~M. \& {Hage}, J.~I. 1990, \apj, 361, 260

\bibitem[{{Gustafson}(1994)}]{1994AREPS..22..553G}
{Gustafson}, B.~A.~S. 1994, Annual Review of Earth and Planetary Sciences, 22,
  553

\bibitem[{{Heiles}(2000)}]{2000AJ....119..923H}
{Heiles}, C. 2000, \aj, 119, 923

\bibitem[{{Henyey} \& {Greenstein}(1941)}]{1941ApJ....93...70H}
{Henyey}, L.~G. \& {Greenstein}, J.~L. 1941, \apj, 93, 70

\bibitem[{{Hines} {et~al.}(2000){Hines}, {Schmidt}, \&
  {Schneider}}]{2000PASP..112..983H}
{Hines}, D.~C., {Schmidt}, G.~D., \& {Schneider}, G. 2000, \pasp, 112, 983

\bibitem[{{Hong}(1985)}]{1985A&A...146...67H}
{Hong}, S.~S. 1985, \aap, 146, 67

\bibitem[{{Iat{\`i}} {et~al.}(2004){Iat{\`i}}, {Giusto}, {Saija}, {Borghese},
  {Denti}, {Cecchi-Pestellini}, \& {Aiello}}]{2004ApJ...615..286I}
{Iat{\`i}}, M.~A., {Giusto}, A., {Saija}, R., {Borghese}, F., {Denti}, P.,
  {Cecchi-Pestellini}, C., \& {Aiello}, S. 2004, \apj, 615, 286

\bibitem[{{Jackson}(1962)}]{1962clel.book.....J}
{Jackson}, J.~D. 1962, {Classical Electrodynamics} (Classical Electrodynamics,
  New York: Wiley, 1962)

\bibitem[{{Kalas} {et~al.}(2005){Kalas}, {Graham}, \&
  {Clampin}}]{2005Natur.435.1067K}
{Kalas}, P., {Graham}, J.~R., \& {Clampin}, M. 2005, \nat, 435, 1067

\bibitem[{{Kalas} {et~al.}(2004){Kalas}, {Liu}, \&
  {Matthews}}]{2004Sci...303.1990K}
{Kalas}, P., {Liu}, M.~C., \& {Matthews}, B.~C. 2004, Science, 303, 1990

\bibitem[{{Kamiuto} {et~al.}(1993){Kamiuto}, {Saitoh}, \&
  {Tokita}}]{1993JQWIISpectroscRadiarTransfer50}
{Kamiuto}, K., {Saitoh}, S., \& {Tokita}, Y. 1993, J. Quant. Spectrosc. Radiar.
  Transfer, 50, 293

\bibitem[{{Kimura} {et~al.}(2006){Kimura}, {Kolokolova}, \&
  {Mann}}]{2006A&A...449.1243K}
{Kimura}, H., {Kolokolova}, L., \& {Mann}, I. 2006, \aap, 449, 1243

\bibitem[{{Kiselev} \& {Velichko}(1998)}]{1998Icar..133..286K}
{Kiselev}, N.~N. \& {Velichko}, F.~P. 1998, Icarus, 133, 286

\bibitem[{{Kozhurina-Platais} \& {Biretta}(2004)}]{2004kozhurina}
{Kozhurina-Platais}, V. \& {Biretta}, J. 2004, {ACS Polarization
  Calibration---III. Astrometry of Polarized Filters: ACS 2004-11} (Baltimore:
  Space Telescope Science Institute)

\bibitem[{{Krist} {et~al.}(2005){Krist}, {Ardila}, {Golimowski}, {Clampin},
  {Ford}, {Illingworth}, {Hartig}, {Bartko}, {Ben{\'{\i}}tez}, {Blakeslee},
  {Bouwens}, {Bradley}, {Broadhurst}, {Brown}, {Burrows}, {Cheng}, {Cross},
  {Demarco}, {Feldman}, {Franx}, {Goto}, {Gronwall}, {Holden}, {Homeier},
  {Infante}, {Kimble}, {Lesser}, {Martel}, {Mei}, {Menanteau}, {Meurer},
  {Miley}, {Motta}, {Postman}, {Rosati}, {Sirianni}, {Sparks}, {Tran},
  {Tsvetanov}, {White}, \& {Zheng}}]{2005AJ....129.1008K}
{Krist}, J.~E., {Ardila}, D.~R., {Golimowski}, D.~A., {Clampin}, M., {Ford},
  H.~C., {Illingworth}, G.~D., {Hartig}, G.~F., {Bartko}, F., {Ben{\'{\i}}tez},
  N., {Blakeslee}, J.~P., {Bouwens}, R.~J., {Bradley}, L.~D., {Broadhurst},
  T.~J., {Brown}, R.~A., {Burrows}, C.~J., {Cheng}, E.~S., {Cross}, N.~J.~G.,
  {Demarco}, R., {Feldman}, P.~D., {Franx}, M., {Goto}, T., {Gronwall}, C.,
  {Holden}, B., {Homeier}, N., {Infante}, L., {Kimble}, R.~A., {Lesser}, M.~P.,
  {Martel}, A.~R., {Mei}, S., {Menanteau}, F., {Meurer}, G.~R., {Miley}, G.~K.,
  {Motta}, V., {Postman}, M., {Rosati}, P., {Sirianni}, M., {Sparks}, W.~B.,
  {Tran}, H.~D., {Tsvetanov}, Z.~I., {White}, R.~L., \& {Zheng}, W. 2005, \aj,
  129, 1008

\bibitem[{{Krivova} {et~al.}(2000){Krivova}, {Krivov}, \&
  {Mann}}]{2000ApJ...539..424K}
{Krivova}, N.~A., {Krivov}, A.~V., \& {Mann}, I. 2000, \apj, 539, 424

\bibitem[{{Kruegel}(2003)}]{2003pid..book.....K}
{Kruegel}, E. 2003, {The physics of interstellar dust} (The physics of
  interstellar dust, by Endrik Kruegel.~IoP Series in astronomy and
  astrophysics, ISBN 0750308613.~Bristol, UK: The Institute of Physics, 2003.)

\bibitem[{{Kundu} {et~al.}(1987){Kundu}, {Jackson}, {White}, \&
  {Melozzi}}]{1987ApJ...312..822K}
{Kundu}, M.~R., {Jackson}, P.~D., {White}, S.~M., \& {Melozzi}, M. 1987, \apj,
  312, 822

\bibitem[{{Larwood} \& {Kalas}(2001)}]{2001MNRAS.323..402L}
{Larwood}, J.~D. \& {Kalas}, P.~G. 2001, \mnras, 323, 402

\bibitem[{{Li} \& {Greenberg}(1997)}]{1997A&A...323..566L}
{Li}, A. \& {Greenberg}, J.~M. 1997, \aap, 323, 566

\bibitem[{{Liu}(2004)}]{2004Sci...305.1442L}
{Liu}, M.~C. 2004, Science, 305, 1442

\bibitem[{{Liu} {et~al.}(2004){Liu}, {Matthews}, {Williams}, \&
  {Kalas}}]{2004ApJ...608..526L}
{Liu}, M.~C., {Matthews}, B.~C., {Williams}, J.~P., \& {Kalas}, P.~G. 2004,
  \apj, 608, 526

\bibitem[{{Metchev} {et~al.}(2005){Metchev}, {Eisner}, {Hillenbrand}, \&
  {Wolf}}]{2005ApJ...622..451M}
{Metchev}, S.~A., {Eisner}, J.~A., {Hillenbrand}, L.~A., \& {Wolf}, S. 2005,
  \apj, 622, 451

\bibitem[{{Pavlovsky}(2006)}]{2006pavlovsky}
{Pavlovsky}, C. 2006, {ACS Data Handbook, Version 5.0} (Baltimore: Space
  Telescope Science Institute)

\bibitem[{{Petrova} {et~al.}(2000){Petrova}, {Jockers}, \&
  {Kiselev}}]{2000Icar..148..526P}
{Petrova}, E.~V., {Jockers}, K., \& {Kiselev}, N.~N. 2000, Icarus, 148, 526

\bibitem[{{Pettersen} \& {Hsu}(1981)}]{1981ApJ...247.1013P}
{Pettersen}, B.~R. \& {Hsu}, J.-C. 1981, \apj, 247, 1013

\bibitem[{{Saar} {et~al.}(1994){Saar}, {Martens}, {Huovelin}, \&
  {Linnaluoto}}]{1994A&A...286..194S}
{Saar}, S.~H., {Martens}, P.~C.~H., {Huovelin}, J., \& {Linnaluoto}, S. 1994,
  \aap, 286, 194

\bibitem[{{Schmidt} {et~al.}(1992){Schmidt}, {Elston}, \&
  {Lupie}}]{1992AJ....104.1563S}
{Schmidt}, G.~D., {Elston}, R., \& {Lupie}, O.~L. 1992, \aj, 104, 1563

\bibitem[{{Strubbe} \& {Chiang}(2005)}]{2005astro.ph.10527S}
{Strubbe}, L.~E. \& {Chiang}, E.~I. 2005, ArXiv Astrophysics e-prints

\bibitem[{{Tamura} {et~al.}(2006){Tamura}, {Fukagawa}, {Kimura}, {Yamamoto},
  {Suto}, \& {Abe}}]{2006ApJ...641.1172T}
{Tamura}, M., {Fukagawa}, M., {Kimura}, H., {Yamamoto}, T., {Suto}, H., \&
  {Abe}, L. 2006, \apj, 641, 1172

\bibitem[{{van de Hulst}(1981)}]{1981lssp.book.....V}
{van de Hulst}, H.~C. 1981, {Light scattering by small particles} (New York:
  Dover, 1981)

\bibitem[{{Voshchinnikov} \& {Kr{\"u}gel}(1999)}]{1999A&A...352..508V}
{Voshchinnikov}, N.~V. \& {Kr{\"u}gel}, E. 1999, \aap, 352, 508

\bibitem[{{Wolstencroft} {et~al.}(1995){Wolstencroft}, {Scarrott}, \&
  {Gledhill}}]{1995Ap&SS.224..395W}
{Wolstencroft}, R.~D., {Scarrott}, S.~M., \& {Gledhill}, T.~M. 1995, \apss,
  224, 395

\bibitem[{{Wurm} \& {Blum}(1998)}]{1998Icar..132..125W}
{Wurm}, G. \& {Blum}, J. 1998, Icarus, 132, 125

\bibitem[{{Zuckerman} {et~al.}(2001){Zuckerman}, {Song}, {Bessell}, \&
  {Webb}}]{2001ApJ...562L..87Z}
{Zuckerman}, B., {Song}, I., {Bessell}, M.~S., \& {Webb}, R.~A. 2001, \apjl,
  562, L87

\end{thebibliography}
\end{document}